\documentclass[intlimits,twoside,a4paper]{article}
\usepackage{amsmath,amssymb}
\usepackage{graphicx}
\usepackage{wrapfig}

\usepackage[T2A]{fontenc}
\usepackage[cp1251]{inputenc}

\usepackage[]{cmpj2}

\issue{2016}{19}{1}{13601}
\doinumber{10.5488/CMP.19.13601}

\title[Adsorption from binary solutions on chemically bonded phases]%
{Adsorption from binary solutions on chemically bonded phases\thanks{This work is dedicated to Professor Stefan Soko\l owski.  We are deeply grateful to him for the longstanding and  fruitful cooperation. \protect}}
\author[M. Bor\'owko, T. Staszewski]{M. Bor\'owko, T. Staszewski}
\address{Department for the Modeling of Physico-Chemical Processes,
Maria Curie-Sk{\l}odowska University,
20-031 Lublin, Poland}

\date{Received September 18, 2015, in final form November 10, 2015}
\authorcopyright{M. Bor\'owko, T. Staszewski, 2016}

\begin{document}

\newcommand{\mtau}{\mbox{\boldmath$\tau$}}
\newcommand{\mqu}{\mbox{\boldmath$q$}}
\newcommand{\mbr}{\mbox{\boldmath$r$}}
\newcommand{\mbR}{\mbox{\boldmath$R$}}
\newcommand{\mbu}{\mbox{\boldmath$u$}}

\maketitle
\begin{abstract}
We use density functional theory to investigate adsorption of liquid mixtures on solid surfaces modified with end-grafted chains. The chains are modelled as freely joined spheres. The fluid molecules are spherical. All spherical species interact via the Lennard-Jones (12-6) potential. The Lennard-Jones (9-3) potential describes interactions of solvent molecules with the substrate. We study the relative excess adsorption isotherms, the structure of surface layer and its composition. The impact of the following parameters on adsorption is discussed: the grafting density, the grafted chain length, interactions of solvent molecules with grafted chains and with the substrate, and the presence of active groups in grafted chains. The theoretical results are consistent with experimental observations.
\pacs 68.47.Mn, 61.25.H, 68.47.Pe, 82.35.Gh
\end{abstract}

\section{Introduction}

A great deal of research has focused on adsorption from solutions on solid surfaces modified with end-grafted chains \cite{c1}. Understanding the adsorption equilibrium is of enormous importance for a variety of biological and technological processes. One of them is the reversed-phase liquid chromatography (RPLC). This technique is among the most popular methods for separation of sample components.

Various theoretical methods have been used to study the retention in chromatography with chemically bonded phases. Among these are  lattice-based analytical theories \cite{c2,c3,c4}, self-consistent field methods \cite{c5,c6}, density functional theory \cite{c7,c8} and computer simulations \cite{c9,c10,c11,c12,c13,c14,c15,c16,c17}. The retention is driven by the distribution of solute molecules between the mobile phase  and the stationary phase (alkyl chains tethered to silica surface).
The process presents theoretical challenges due to its complexity, and thus numerous problems related to its mechanism are still unsettled.

The aim of the chromatographic analysis is to achieve the elution of all sample components in reasonable time and with a satisfactory selectivity of the separation.  To optimize this process, one commonly uses  a mixture of two solvents. Changing the composition of the mobile phase we change the system selectivity. Much effort has been  directed toward the theoretical prediction of the solution retention as a function of the mobile phase composition \cite{c2,c18,c19,c20}.

The retention strongly depends on adsorption of solvents at the stationary phase \cite{c2}. For this reason, numerous experimental data referring to adsorption of binary solvents commonly used in chromatography have been recently published  \cite{c20,c21,c22,c23,c24,c25,c26}. However, in the literature one can find only a few theoretical articles connected with adsorption from solutions on chemically bonded phases. The bonded phases are usually treated as `usual' adsorbents, and the theory of adsorption from solutions on solid surfaces is employed to interpret experimental data. The liquid part of a system is formally divided into the surface phase and the bulk phase. The location of the dividing surface is rather arbitrary and several methods for its estimation have been considered \cite{c13,c20}. Gritti and Gouichon \cite{c20} have used the so-called bi-Langmuir equation for excess adsorption isotherms. They assumed that the adsorbent was composed of two patches, one representing the surface covered with grafted chains and the second corresponding to  the bare surface of the solid. Their approach has been used to study adsorption  on different bonded phases \cite{c20,c21,c24}. This phenomenological theory does not give any insight into the structure of the chain layer. Quite recently,  the density functional theory has been used to study adsorption from solutions on the polymer-tethered surfaces \cite{c27,c28,c29,c30}. In this model, the penetration of the solvent molecules into the chain layer is allowed. The densities of all species gradually change with the distance of the solid surface and tend to their bulk values.

In this work, we present the results of the density functional study of the competitive adsorption from binary solutions on chemically bonded phases.
We show how the selected parameters affect the composition of the liquid inside the grafted chain layer. Hitherto, this problem has  not been  analyzed. The local changes in the mixture composition can influence the chromatographic separation.  We also consider  relative excess adsorption isotherms  for the  model systems investigated.  We discuss the impact of such parameters as: the grafting density, the grafted chain length and interactions of solvents with grafted chains and with the substrate, the presence of active groups in the chains. We want to find general trends rather than to approximate experimental data for concrete systems. Our conclusions  are  consistent with the results  of experiments and computer simulations found in the literature. Short grafted chains are used in popular stationary phases. Therefore, we concentrated on grafted oligomers.

The article is organized as follows. In the next section, we describe the model and the basic aspects of the theoretical approach. The results are presented and analyzed in section~3. Finally, we summarize the conclusions.

\section{Model and theory}

We study adsorption from a binary solution on a surface modified with end-grafted chains. We employ the computational method used in our previous papers
concerning this problem \cite{c27,c28,c29,c30}. Therefore, we discuss here only the most important aspects of the model.
The method was originally proposed by Yu and Wu \cite{c31,c32,c33}. Numerous research groups used the density functional theory to study the systems involving either free  \cite{c30a,c30b,c30c,c30d} or grafted chains  \cite{c30e,c30f,c30g,c30h, c30i}.
We treat the system as a ternary mixture in contact with an impenetrable wall. The mixture consists of tethered polymers (P) and molecules of solvents (labeled as 1 and 2). The grafted polymers (P) are chains of $M$ freely jointed segments. The connectivity  of a given chain is enforced by the bonding potential
\begin{equation} \label{eq:1}
\exp [-\beta V_{B}({\bf R})]=\prod_{i=1}^{M-1}\delta (|{\bf r}_{i+1}-%
{\bf r}_{i}|-\sigma^{(\text{P})})/4\pi (\sigma^{(\text{P})})^{2},
\end{equation}
where ${\bf R}_k\equiv (\mathbf{r}_{1},\mathbf{r}_{2},\ldots ,\mathbf{r}%
_{M})$ is the vector specifying the positions of all segments,  $\sigma^{(\text{P})}$ is the segment diameter, the symbol $\delta$ denotes
the Dirac function, $\sigma^{(\text{P})}$ is the polymer segment diameter
and $\beta^{-1}=k_\text{B}T$.

The first segment of the chain is bonded with the surface by the potential
\begin{equation} \label{eq:2}
\exp\left[-\beta v_{\text{s}1}^{(\text{P})}(z)\right]=C \delta (z-\sigma^{(\text{P})}/2),
\end{equation}
where $z$ is a distance from the surface and $C$ is a constant. The surface-binding segments are located at the distance $z=\sigma^{(\text{P})}/2$ from the wall. These segments cannot leave the surface but they can move within the $xy$-plane.

All of the remaining segments, $i=2, 3, \ldots, M$ are neutral with respect to the surface and  interact with the substrate via the hard wall potential
\begin{equation} \label{eq:3}
v^{(k)}=\left\{
\begin{array}{ll}
\infty, & \ \ \ z < \sigma^{(\text{P})}/2, \\
0, & \ \ \ \text{otherwise},
\end{array}
\right.
\end{equation}
where   $v^{(\text{P})}=v^{(\text{P})}_{\text{s}i}$ ($i \geqslant 2$).

The solvent molecules are attracted by the surface, according to the Lennard-Jones (9--3) potential
\begin{equation} \label{eq:4}
v^{(k)}=4 \bar \varepsilon^{(k)}_\text{s} \left[ (z_0^{(k)}/z)^{9}-(z_0^{(k)}/z)^3 \right],
\end{equation}
where $\bar \varepsilon^{(k)}_\text{s}$ characterizes the strength of interactions between the $k$th solvent and the wall ($k=1,2$) and $z_0=\sigma^{(k)}/2$. We also consider the solvents neutral with respect to the substrate, interacting with the surface via the hard-wall potential [equation~(\ref{eq:3})].

The chain segments and fluid molecules interact
via Lennard-Jones potential (12--6)
\begin{equation} \label{eq:5}
u^{(kl)}=\left\{
\begin{array}{ll}
4 \bar \varepsilon^{(kl)}\left[(\sigma^{(kl)}/r)^{12}-(\sigma^{(kl)}/r)^6\right],& \ \ \ r < r_\text{cut}^{(kl)}, \\
0, & \ \ \ \text{otherwise},
\end{array}
\right.
\end{equation}
where $\bar \varepsilon^{(kl)}$ is the parameter characterizing interactions between species $k$ and $l$, $\sigma^{(kl)}=0.5(\sigma^{(k)}+\sigma^{(l)})$ for $k,l=1,2,\text{P}$; $r$ is
the distance between the interacting spheres, $r_\text{cut}^{(kl)}$ is the cutoff distance.
In this work $r_\text{cut}^{(kl)}=3\sigma^{(kl)}$.

We assume that the grafting density, $\rho_\text{P}=N_\text{P}/A_\text{s}$, is fixed where $N_\text{P}$ is the number
of grafted chains and  $A_\text{s}$ denotes the area of the surface.

The theory is constructed in terms of  the local densities of spherical molecules, $\rho^{(k)}$ ($k=1,2$)
and the local density of segments of the grafted chains
\begin{equation}  \label{eq:6}
\rho _\text{s}^{(\text{P})}({\bf r})=\sum_{i=1}^{M}\rho _{\text{s},i}^{(\text{P})}({\bf r}%
)=\sum_{i=1}^{M}\int \rd \mathbf{R}\delta (\mathbf{r}-{\bf r}%
_i)\rho^{(\text{P})}({\bf R})\;,
\end{equation}
where $\rho^{(\text{P})}$ is the local density of the chains and $\rho^{(\text{P})}_{\text{s},i}$ is
the density of $i$-th segments.

Using the procedure prosed by Yu and Wu \cite{c31,c32,c33} we calculate the density profiles of all components. As usually, the free-energy functional is expressed as the sum $F=F_\text{id}+F_\text{hs}+F_\text{c}+F_\text{att}$.
The free energy of
an ideal gas, $F_\text{id}$, is known exactly \cite{c34}. The excess free energy
following from hard-sphere interactions, $F_\text{hs}$, is calculated from a modified version \cite{c32}
of the fundamental measure theory of Rosenfeld \cite{c35}.
The chain connectivity contribution, $F_\text{c}$, follows from
the first-order perturbation theory of Wertheim \cite{c36}.  A reader can find all
necessary expressions in reference \cite{c37} [equations (7), (10) and (12), respectively]. The attractive interactions
between spherical species are expressed using mean-field approximation
\begin{eqnarray} \label{eq:7}
F_\text{att} &=& \frac{1} {2} \sum_{k=\text{P},1,2} \int \rd {\bf r}_1 \rd {\bf r}_2
\rho_\text{s}^{(k)}({\bf r}_1) \rho_\text{s}^{(k)}({\bf r}_2) u^{(kk)}_\text{att}({\bf r}_{12})\nonumber\\
&&+ \sum_{\begin{smallmatrix}k,l=\text{P},1,2;\\ k<l\end{smallmatrix}} \int \rd {\bf r}_1 \rd {\bf r}_2 \rho_\text{s}^{(k)}({\bf r}_1)
\rho_\text{s}^{(l)}({\bf r}_2) u^{(kl)}_\text{att}({\bf r}_{12}),
\end{eqnarray}
where $u^{(kl)}_\text{att}$ is the attractive part of Lennard-Jones potential following
from the Weeks-Chandler-Ander\-son scheme \cite{c38}
\begin{equation} \label{eq:8}
u^{(kl)}_\text{att}(r)=\left\{
\begin{array}{ll}
-\varepsilon^{(kl)}, &  r <  2^{1/6} \sigma^{(kl)}, \\
u^{(kl)}(r),& r \geqslant 2^{1/6} \sigma^{(kl)},
\end{array}
\right.
\end{equation}
and  $\rho^{(1)}_\text{s}=\rho^{(1)}$ and $\rho^{(2)}_\text{s}=\rho^{(2)}$.

The grafting density, $\rho_\text{P}$, is given by
\begin{equation} \label{eq:9}
\int_0 \rd{z}\rho_{\text{s},i}^{(\text{P})}(z) =\rho_\text{P}.
\end{equation}

The  equilibrium density profiles are obtained by minimizing the thermodynamic potential
\begin{eqnarray} \label{eq:10}
{ Y} &=& F+\int \rd{\bf R}\rho^{(\text{P})}({\bf R})v^{(\text{P})}({\bf R})\nonumber\\
&&+ \sum_{k=1,2} \int \rd{\bf r}_k\rho^{(k)}({\bf r}_k)(v^{(k)}({\bf r}_k)-\mu^{(k)})
\end{eqnarray}
under the constraint (\ref{eq:9}).   In the above, $\mu^{(k)}$ denotes the chemical potential of the component
$k$ ($k=1, 2$).  In the considered model, the density distributions vary only with the distance
from the surface, $\rho^{(k)}(\mathbf{r})=\rho^{(k)}(z)$, $k=\text{P},1,2$. Minimization of the thermodynamic potential
leads to a set of the Euler-Lagrange equations (equation (17) in reference \cite{c37}) which can be solved numerically. Knowing the equilibrium distributions of all components we can calculate different quantities characterizing the adsorption and the structure of the surface layer.

In the case of liquid mixtures, different species compete for room in the surface region, molecules of a given component are displaced by molecules of the other component. As a consequence, the composition of a solution considerably depends on the distance from the wall. To characterize the composition of the solution we define the local volume fraction of the $k$th component ($k=1,2$)\begin{equation} \label{eq:11}
 x^{(k)}= \frac {\rho^{(k)} (\sigma^{(k)})^3} {\rho^{(1)}
 (\sigma^{(1)})^3 + \rho^{(2)} (\sigma^{(2)})^3}\,, \qquad  k=1,2.
\end{equation}
Notice that $x^{(k)}$ is not the volume fraction in the whole system but only in the liquid.

For a competitive adsorption from a solution, the relative excess adsorption isotherms are usually
used \cite{c20,c24,c39,c40}
\begin{equation} \label{eq:12}
N^\text{e}_k = \int \rd z \left [ x^{(k)}(z)-x_{b}^{(k)} \right ],
\end{equation}
and $x^{(k)}_\text{b}$ is the volume fraction in the bulk mixture. Obviously, $N^\text{e}_1=-N^\text{e}_2$.
The relative excess $N^\text{e}_k$  can be directly
compared with experimental data because it is proportional to the difference in the bulk solution composition after and before the adsorption $\Delta x=x^{(k)}_\text{b}-x_0^{(k)}$.

The theory involves numerous molecular parameters. The chain segments and solvent molecules can have different sizes. The bonded phase is characterized by the length of chains, $M$, and by the grafting density, $\rho_\text{P}$. Interactions between molecules of different components are characterized by the energy parameters $\varepsilon^{(kl)}$  ($k,l=\text{P},1,2$) while the parameter $\varepsilon^{(k)}_\text{s}$ ($k=1,2$) describes interactions with the substrate. Moreover, the densities of components 1 and 2 ($\rho^{(1)}_\text{b}$, $\rho^{(2)}_\text{b}$) in the bulk solution should be specified.

We express the energy parameters  in units of thermal energy $k_\text{B}T$. In this order we define dimensionless energy parameters
$\varepsilon^{(kl)}=\bar \varepsilon^{(kl)}/k_\text{B}T$ and $\varepsilon_\text{s}^{(k)}=\bar \varepsilon_\text{s}^{(k)}/k_\text{B}T$.
To compare our model with other theories we introduce the Flory-Huggins
type parameters $\chi^{(kl)}=-[\varepsilon^{(kl)}-0.5(\varepsilon^{(kk)}+\varepsilon^{(ll)})]$.
The parameter $\chi^{(12)}$ characterizes the nature of the bulk fluid.
The parameter $\chi^{(\text{P}k)}$ describes the compatibility of the $k$-th component with respect to the chain segments.

\section{Results and discussion}

An adsorption equilibrium is the result of a complex interplay between two basic factors, i.e., the attractive interactions in the systems and the repulsion in the film built of grafted chains.  The bonded chains can act as a barrier for the solvent molecules that tend to the substrate. On the other hand, they can provide additional `adsorption sites'. The adsorption process depends in a very complicated way on the properties of all components of the system: the solid surface, grafted chains and the solvents.

The aim of our study is to show how the selected parameters affect adsorption  and the composition of the surface layer. We present here the results of the systematic model calculations carried out for different combinations of the parameters. We try to mimic a polar adsorbent (e.g., silica gel) modified with tethered alkyl-like chains in contact with the mixture of solvents.

As already mentioned, the system under study comprises several parameters. In order to reduce their number to
 a minimum we assume that   diameters of all
the segments, as well as of all the free molecules are the same,  $\sigma^{(\text{P})}=\sigma^{(1)}=\sigma^{(2)}=1$.  Moreover, the calculations have been carried out by fixing the following parameters: $\varepsilon^{(\text{PP})}=\varepsilon^{(\text{P}2)}=\varepsilon^{(11)}
=\varepsilon^{(22)}=\varepsilon^{(12)}=1$ and $\varepsilon^{(1)}_\text{s}=1$. The total bulk density of the fluid is $\rho^{(\text{F})}=\rho^{(1)}+\rho^{(2)}=0.8$. Notice that  $\chi^{(\text{P}2)}=0$.

In this work the component 1 mimics an organic solvent that has a high affinity to tethered chains but a relatively weak affinity to the polar surface. The component 2, however, can be treated either as water or as an organic solvent, %
depending on the value of the parameter $\varepsilon^{(2)}_\text{s}$.   We assumed that interactions of `polar' surface with the solvent 2 are  considerably stronger than the interactions with the solvent 1 ($\varepsilon^{(2)}_\text{s} >\varepsilon^{(1)}_\text{s}$). We focus our attention on the role of the modified adsorbent. To simplify the analysis of the results, we assume that the liquid mixture is ideal ($\chi^{(12)}=0$).

\begin{figure}[!t]
\centerline{
\includegraphics[width=0.55\textwidth]{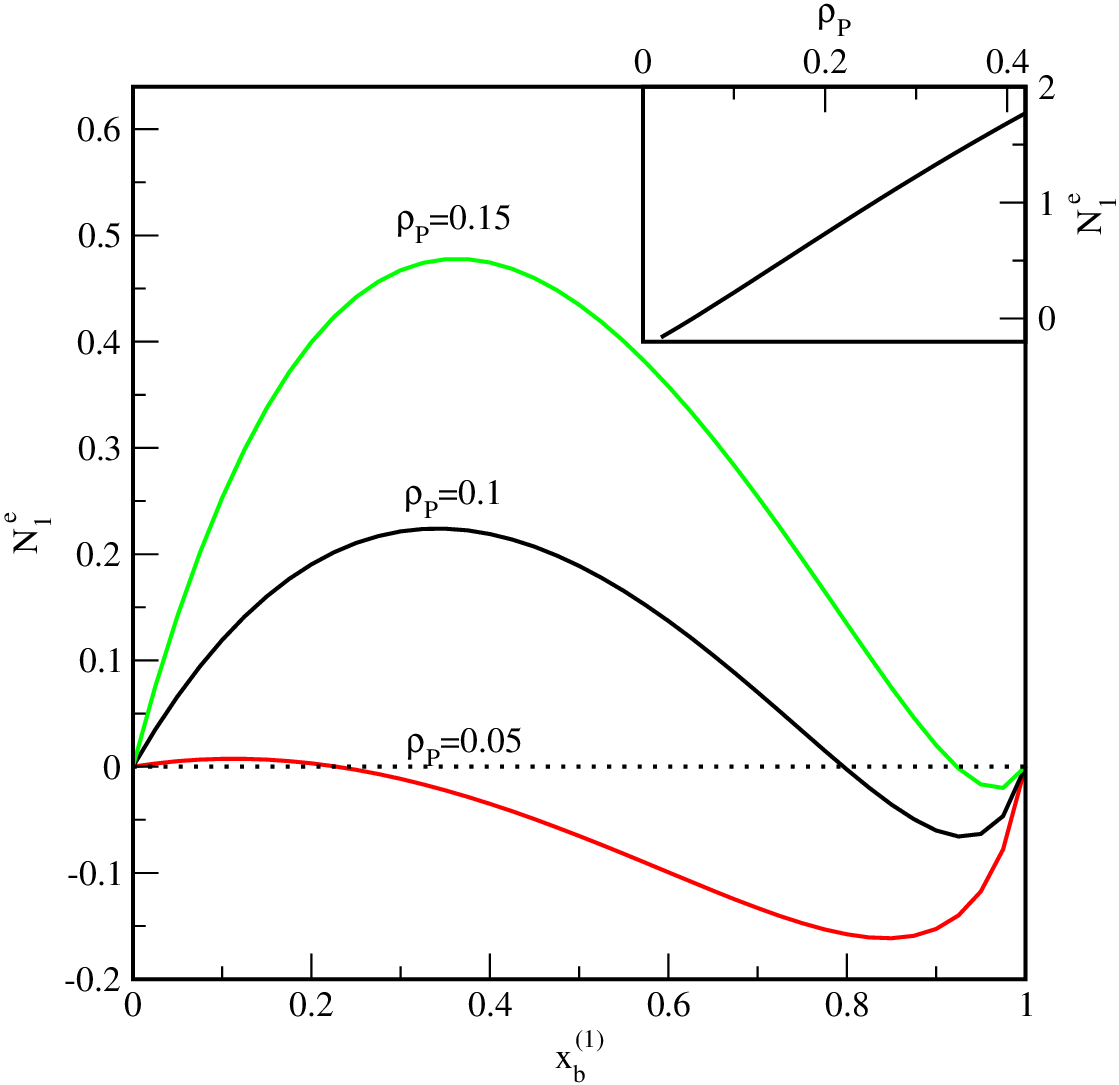}
}
\caption{(Color online) Relative excess adsorption isotherms for different values of the grafting density: $\rho_\text{P}=0.05$, $0.1$, $0.15$. The inset presents the influence of the grafting density on the relative adsorption, $N^\text{e}_1$, for the fixed composition of the bulk solution,  $x^{(1)}_\text{b}=0.35$. Parameters: $M=8$, $\varepsilon^{(\text{P}1)}=1.2$, $\varepsilon_\text{s}^{(2)}=10$. }
\label{fig:1}
\end{figure}

We begin with the discussion of the influence of the grafting density on the adsorption from solutions  on the chemically bonded phases. We consider here the surface layer built of grafted octamers ($M=8$). The relative excess adsorption isotherms obtained for different grafting densities are shown in figure~\ref{fig:1}. The remaining parameters do not vary. In this case, the 1st solvent has only a slightly higher affinity to the grafted chains than the other component: $\varepsilon^{(\text{P}1)}=1.2$ and $\varepsilon^{(\text{P}2)}=1$. However, the interaction of the 1st component with the substrate is much weaker than the interactions of the 2nd solvent: $\varepsilon^{(1)}_\text{s}=1$  but $\varepsilon^{(2)}_\text{s}=10$. One sees that an increase of the grafting density causes a considerable increase of the relative excess adsorption $N^\text{e}_1$.
In the inset, the relation $N^\text{e}_1$ vs $\rho_\text{P}$ is presented for the fixed composition ($x^{(1)}_\text{b}=0.35$). The more grafted are the chains, the  more profitable are the contacts between the molecules 1 and the chain segments. As a consequence, the sorption increases. Such a trend was observed in experimental data \cite{c21,c24,c25}.

Adsorption azeotropy is observed in the considered systems.  At the azeotropic point, $x^{(1)}_\text{b,az}$, the relative excess adsorption equals zero: $N^\text{e}_1(x^{(1)}_\text{b,az})=0$ for $0<x^{(1)}_\text{b,az}<1$. One can say that at the azeotropic point, the composition of the liquid in the surface layer is the same as the composition of the bulk solution. With an increasing grafting density, the azeotropic point shifts toward higher concentrations of the 1st component  For $x^{(1)}>x^{(1)}_\text{b,az}$,  the preferential adsorption of the 2nd solvent is found. The maximum value of the relative excess adsorption of component 1 increases with an increase of grafting density while an opposite relation is observed for the component 2.

Our conclusions agree with the analysis of experimental data \cite{c21,c24,c26}. Gritti et al. \cite{c21} measured the adsorption of acetonitrile, tetrahydrofuran and alcohols from water on end-capped silica. They have shown that an increase of the grafting density causes a rise of the relative excess adsorption of acetonitrile and tetrahydrofuran and shifts the adsorption azeotropic points toward the higher concentration of organic modifiers. In the case of the investigated alcohols, the same trend is observed for dilute solutions and the grafting densities below a certain threshold value. The analysis of the adsorption isotherms obtained for non-end-capped C$_{18}$-bonded phases leads to the same conclusions \cite{c24}.

These results are qualitatively consistent with the predictions of the analytical theory of adsorption on energetically heterogeneous solid surfaces \cite{c42}.  Adsorption azeotropy can be caused either by the nonideality of the solution or by the heterogeneity of the surface. In the case of ideal solutions, as there increases the number of active sites exhibiting a stronger affinity to the molecules 1, the azeotropy point tends to unity \cite{c42}. Gritti and Gouichon \cite{c20} have shown that the chemically bonded phases behave as energetically heterogeneous adsorbents. They assumed that there are two kinds of adsorption sites on the surface: grafted chains and patches of bare substrate. Within the framework of their theory, with an increase of fraction of the surface covered with the chains,  the adsorbent becomes more homogeneous and the azeotropic point tends to unity. In the Gritti-Gouichon \cite{c20} approach, the composition of the surface layer is assumed to be independent of the distance from the wall. There is no such a limitation in our treatment.

\begin{figure}[!t]
\centerline{
\includegraphics[width=0.55\textwidth]{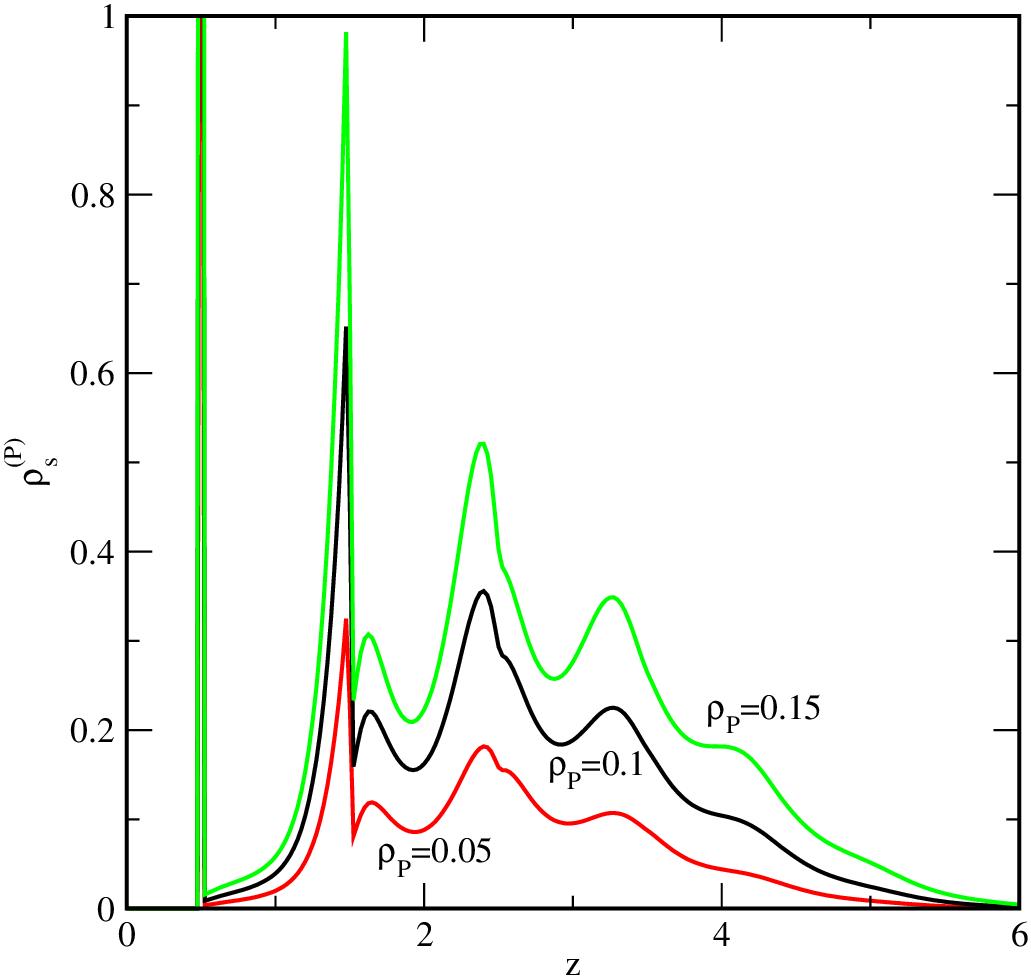}
}
\caption{(Color online) Segment density profiles of grafted chains for different grafting densities, $\rho_\text{P}=0.05$, $0.1$, $0.15$, at $x^{(1)}_\text{b}=0.35$. The remaining parameters are the same as those in figure~\ref{fig:1}.}
\label{fig:2}
\end{figure}

Figure~\ref{fig:2}  shows the segment density profiles of grafted chains  for different grafting densities. The profiles have a typical liquid-like structure with peaks corresponding to successive layers of chain segments near the wall. In the outer region of the bonded phase, the chain density smoothly decreases to zero.
The height of the chain layer considerably increases with an increase of the grafting density. The repulsive forces between the chains enforce them to stretch in the direction perpendicular to the wall.

\begin{figure}[!t]
\centerline{
\includegraphics[width=0.6\textwidth]{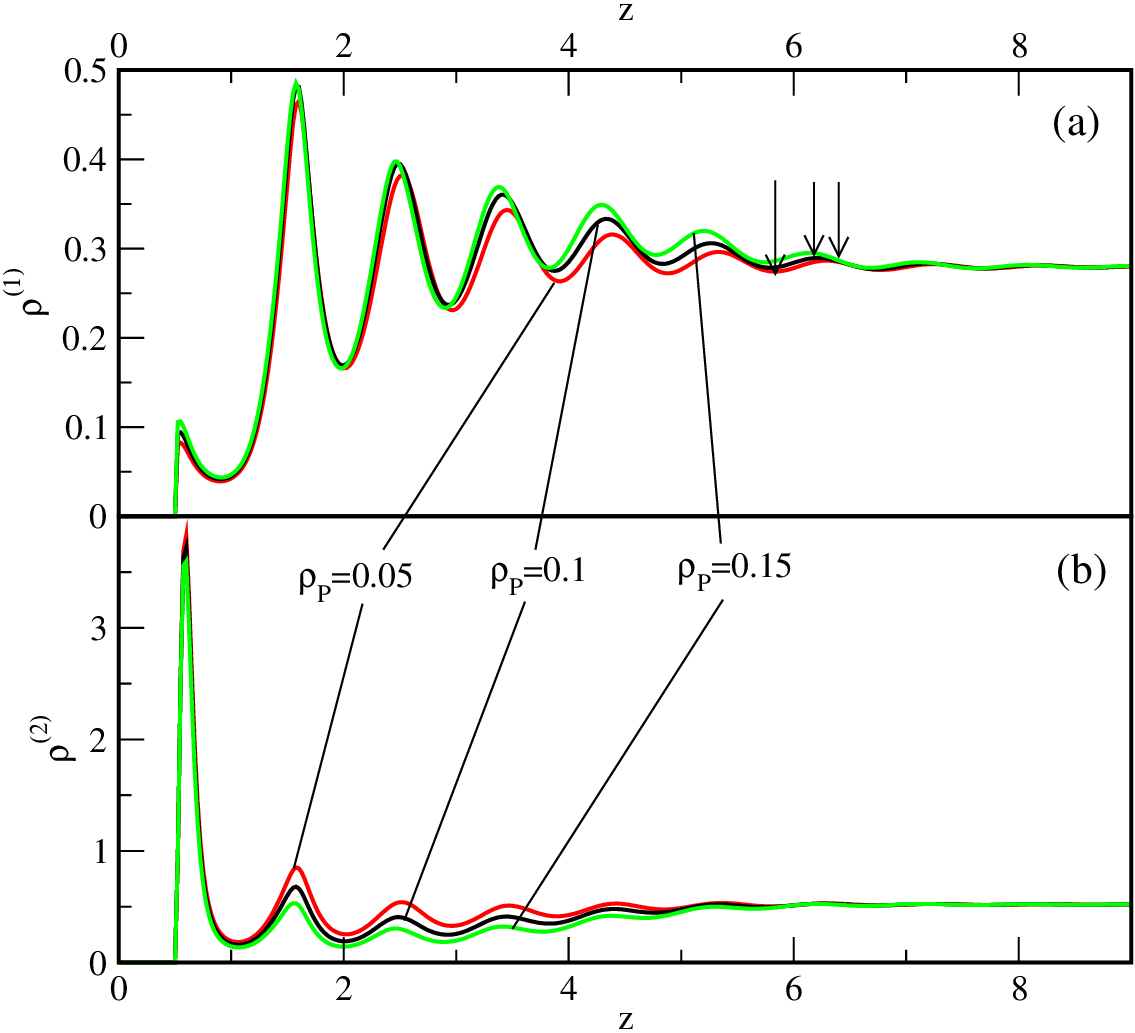}
}
\caption{(Color online) Density profiles of the first solvent (panel a) and the second solvent (panel b) for different grafting densities, $\rho_\text{P}=0.05$, $0.1$, $0.15$, at $x^{(1)}_\text{b}=0.35$. The remaining parameters are the same as those in figure~\ref{fig:1}. Arrows indicate the chain layer edges.}
\label{fig:3}
\end{figure}

The density profiles of the solution components presented in  figure~\ref{fig:3} provide data on the solvent distribution in the surface layer. The arrows show the brush edges for different grafting densities. The  results have been obtained  for the average concentration of the 1st component, $x^{(1)}_\text{b}=0.35$. In the considered system, both solvents have relatively high affinity to the grafted chains, while the grafting density is rather low. Therefore, all fluid molecules penetrate the chain layer. There are several peaks at the local density profiles of both components. The positions of these peaks correspond to the maxima observed at the density profiles of the chains. The solvent molecules `stick' to the grafted chains. Due to strong interactions with the solid surface, the molecules 2 accumulate on the wall and there is a significant peak near the substrate. The structure of the middle part of the surface layer considerably depends on the grafting density. With an increasing number of grafting chains, the density of the component 1, $\rho^{(1)}$, increases while the density of the other solvent decreases. In the outer region of the surface layer, the component densities gradually tends to their bulk values.  Note that the molecules 1  accumulate not only within the brush but also atop the bonded phase.

\begin{figure}[!b]
\centerline{
\includegraphics[width=0.65\textwidth]{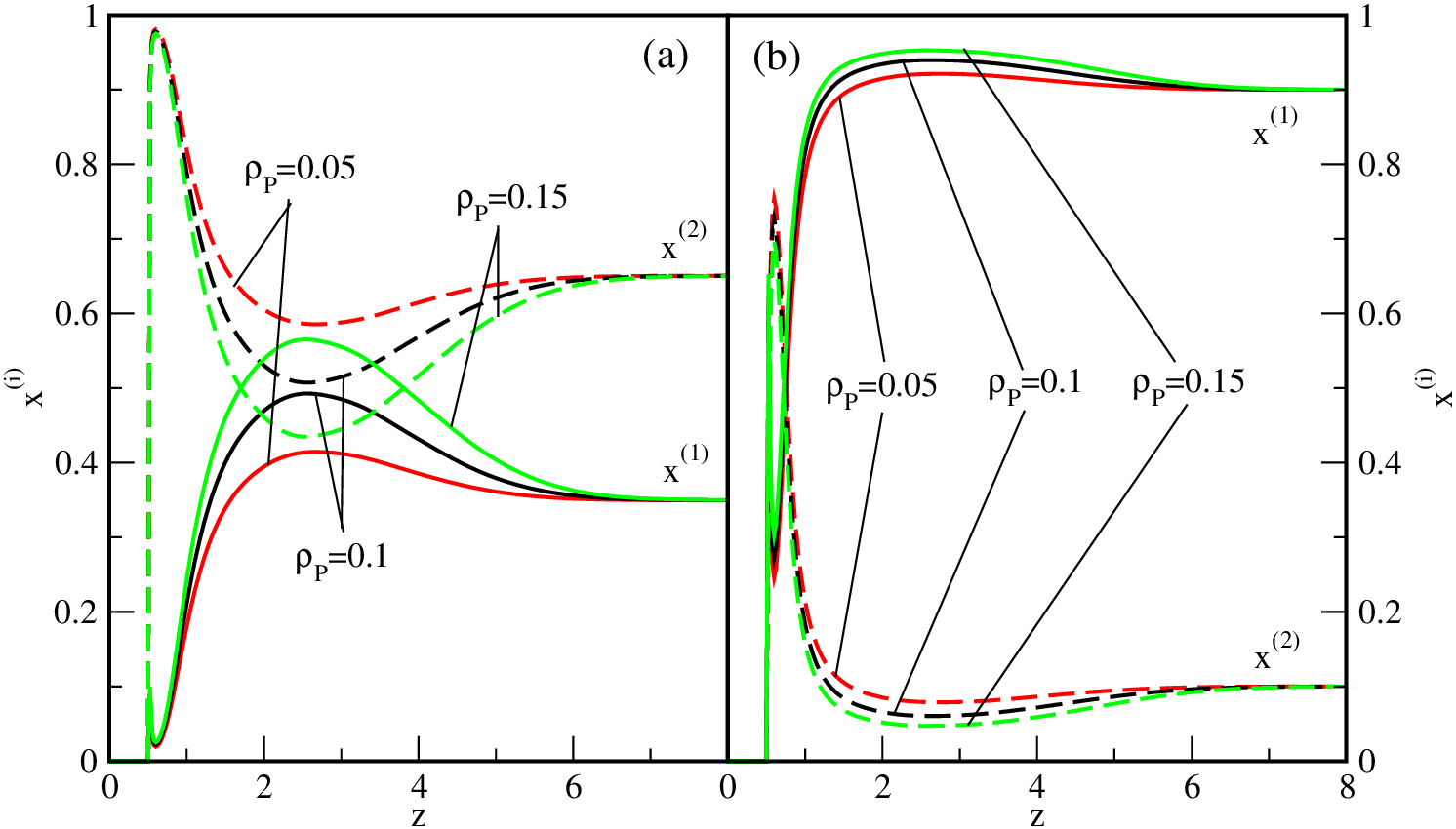}
}
\caption{(Color online) Profiles of the solution composition for different values of the grafting density, $\rho_\text{P}=0.05$, $0.1$, $0.15$, at fixed compositions of the bulk solution: $x^{(1)}_\text{b}=0.35$ (panel a) and  $x^{(1)}_\text{b}=0.90$. Solid (dashed) lines correspond to the local mole fraction of the 1st (2nd) component. The remaining parameters are the same as those in figure~\ref{fig:1}.}
\label{fig:4}
\end{figure}

It is instructive to analyze the local composition of the surface layer. The local mole fractions of liquid mixture  are depicted in figure~\ref{fig:4} for $x^{(1)}_\text{b}=0.35$ (a) and $x^{(1)}_\text{b}=0.9$ (b).  For a given value of the grafting density, the mole fraction $x^{(1)}$ achieves a deep  minimum near the surface and a  maximum in the middle part of the chain layer.   Close to the wall, the mole fraction of the 1st (2nd) component is always considerably lower (greater) than  in the bulk phase. In the region $1.5<z<6.5$, however, the inverse relation is found,  the mole fraction of the 1st (2nd) component is greater (smaller) than its mole fraction in the bulk solution. There is a significant enrichment of the `organic' solvent 1 within the bonded phase.  With an increase of the grafting density, the mole fraction  $x^{(1)}$ increases ($x^{(2)}$ decreases). These effects are much stronger for the lower mole fraction $x^{(1)}_\text{b}=0.35$. We see that the liquid in the surface region is highly inhomogeneous. Our conclusions are in agreement with the results of molecular dynamic simulations reported by Rafferty et al. \cite{c10, c11}.

 \begin{figure}[!t]
\centerline{
\includegraphics[width=0.65\textwidth]{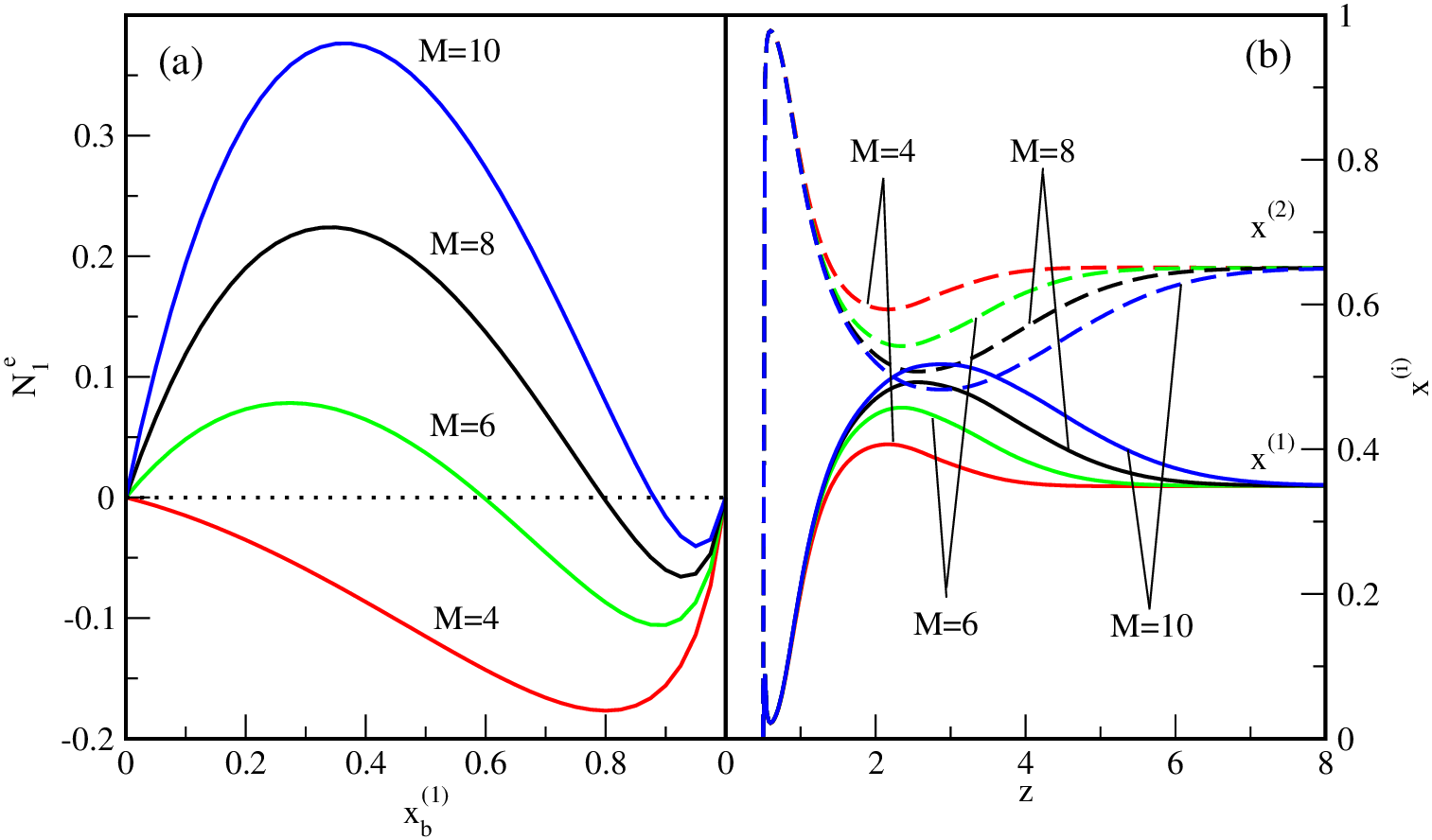}
}
\caption{(Color online) Relative excess adsorption isotherms (panel a) and  profiles of the solution composition at  $x^{(1)}_\text{b}=0.35$ (panel b) for different  lengths of grafted chains: $M=4$, $6$, $8$, $10$.  In panel b, solid (dashed) lines correspond to the local mole fraction of the 1st (2nd) component. Parameters: $\rho_\text{P}=0.1$, $\varepsilon^{(\text{P}1)}=1.2$, $\varepsilon_\text{s}^{(2)}=10$.}
\label{fig:5}
\end{figure}

Another way to moderate the adsorptive properties of the bonded phase is to change the length of the grafted chains.  The examples of the relative excess adsorption isotherms calculated for different chain lengths are shown in figure~\ref{fig:5}~(a). The average value of grafting density was assumed to be $\rho_\text{P}=0.1$. The remaining parameters are the same as those in figure~\ref{fig:1}. Under these conditions, the relative excess adsorption $N^\text{e}_1$ is greater for longer grafted chains. On the surface covered by short chains ($M=4$), the 2nd component is adsorbed preferentially in the whole concentration region ($N^\text{e}_1<0$). For longer chains, adsorption of the 1st solvent is favored at  lower mole fractions, $x^{(1)}_\text{b}$. An increase of the number of chain segments causes a rise of a profitable contact between the molecules 1 and the grafted chains, and thus the relative excess adsorption $N^\text{e}_1$ increases.  The same relation has been found for adsorption of organic solvents from aqueous solutions on the bonded phases with different chain lengths attached to silica  \cite{c22,c23}.  For denser brushes and very long polymers, an impact of chain lengths weakens \cite{c22}.

In figure~\ref{fig:5}~(b),  the composition profiles at $x^{(1)}_\text{b}=0.35$ are presented.  Near the substrate the composition of the liquid is almost independent of the grafted chain length.  However, one sees a significant effect of the chain length on the liquid composition in the remaining part of the surface layer. The mole fraction of the 1st component increases for longer grafted chains. Obviously, the opposite is observed for the mole fraction $x_\text{b}^{(2)}$. Analogous results have been obtained from simulation \cite{c14}.

\begin{figure}[!b]
\centerline{
\includegraphics[width=0.65\textwidth]{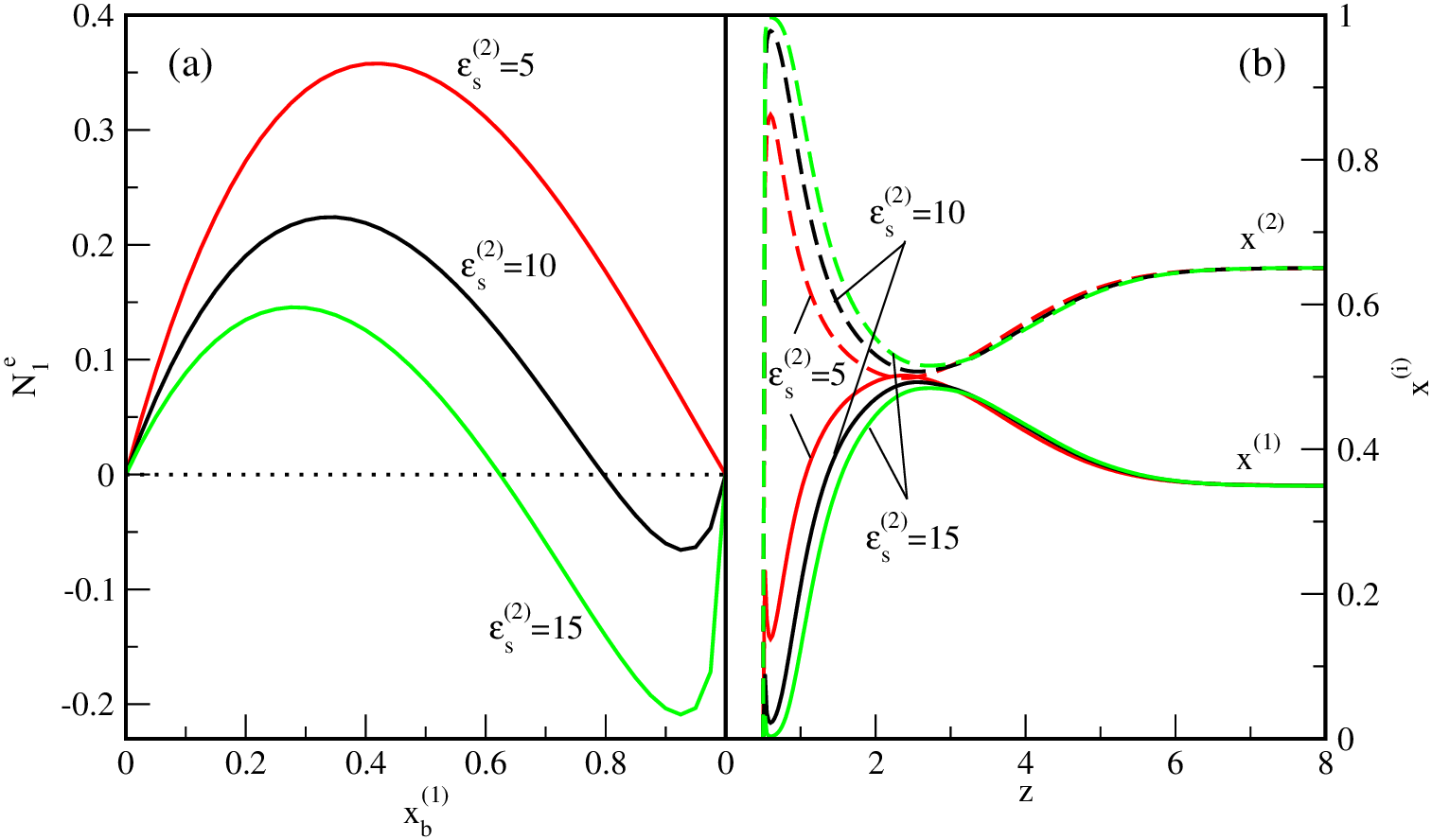}
}
\caption{(Color online) Relative excess adsorption isotherms (panel a) and  profiles of the solution composition at  $x^{(1)}_\text{b}=0.35$ (panel b) for different  values of the energy parameter  $\varepsilon^{(2)}_\text{s}=5$, $10$, $15$.  In panel b, solid (dashed) lines correspond to the local mole fraction of the 1st (2nd) component. Parameters: $\rho_\text{P}=0.1$, $M=8$,  $\varepsilon^{(\text{P}1)}=1.2$.}
\label{fig:6}
\end{figure}

Now we turn to the effects of interactions with the whole adsorbing material, i.e., with the substrate and with the grafted chains. Let us consider the role of interactions with the solid surface.
In figure~\ref{fig:6}~(a), the relative excess isotherms are plotted for a fixed value of energy parameter $\varepsilon^{(1)}_\text{s}=1$ and for three values of the energy parameter $\varepsilon^{(2)}_\text{s}=5, 10, 15$.  We moderate the relative solvent affinity by changing the solvent 2. The relative excess adsorption isotherm reflects the competition in accumulation of components 1 and 2 within the surface layer. The relative affinity of the 1st component to the whole adsorbing material increases  as the difference $\Delta=\varepsilon^{(1)}_\text{s}-\varepsilon^{(2)}_\text{s}$ increases. Indeed, one sees that the preferential adsorption of the component 1 is greater for lower values of the parameter $\varepsilon^{(2)}_\text{s}$. At the same time, the azeotropic point shifts to unity, namely $x^{(1)}_\text{b,az}=0.62$ for $\varepsilon^{(2)}_\text{s}=15$  and $x^{(1)}_\text{b,az}=0.80$ for $\varepsilon^{(2)}_\text{s}=10$. Such a behavior is predicted by the simple theories of adsorption form solutions \cite{c42}. These conclusions are in an agreement with the experimental observations \cite{c24}.

The profiles of the mixture composition are plotted in figure~\ref{fig:6}~(b)  for $x^{(2)}_\text{b}=0.35$. The molecules 2 accumulate at the surface, and the mole fraction $x^{(2)}$ decreases with an increase of the parameter $\varepsilon^{(2)}_\text{s}$ (dashed lines). On the contrary, the mole fraction of the 1st component is considerably lower for higher values of $\varepsilon^{(2)}_\text{s}$. These effects are significant only in the immediate proximity to the wall.

\begin{figure}[!t]
\centerline{
\includegraphics[width=0.65\textwidth]{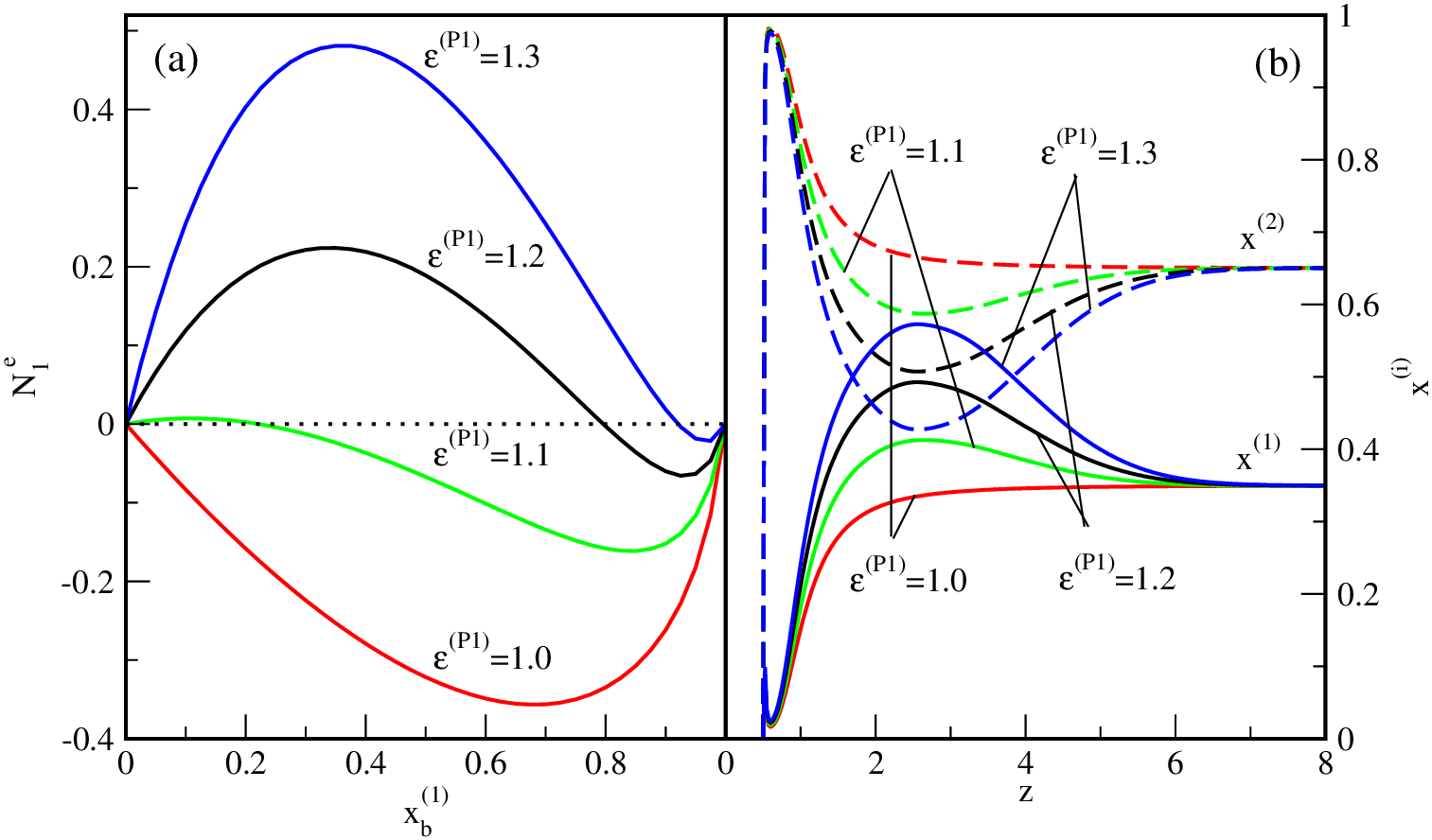}
}
\caption{(Color online) Relative excess adsorption isotherms (panel a) and  profiles of the solution composition at  $x^{(1)}_\text{b}=0.35$ (panel b) for different  values of the energy parameter  $\varepsilon^{(\text{P}1)}=1.0$, $1.1$, $1.2$, $1.3$.  In panel b, solid (dashed) lines correspond to the local mole fraction of the 1st (2nd) component. Parameters: $\rho_\text{P}=0.1$, $M=8$,  $\varepsilon^{(2)}_\text{s}=10$. }
\label{fig:7}
\end{figure}

Adsorption of the 1st component can be also altered by changing the parameter $\varepsilon^{(\text{P}1)}$ [see figure~\ref{fig:7}~(a)].  The relative affinity of the components to the chains is well quantified by the difference $\Delta_\text{P}=\varepsilon^{(\text{P}1)}-\varepsilon^{(\text{P}2)}$. The  energy parameter  $\varepsilon^{(\text{P}2)}=1$. An increase of the parameter  $\varepsilon^{(\text{P}1)}$ leads to a considerable rise of the preferential adsorption of the 1st component.  The attractive interactions of the molecules 1 with segments of the grafted chains markedly affect  the composition of the middle and the outer parts of the surface region. This is clearly demonstrated in figure~\ref{fig:7}~(b).

Adsorptive properties of the bonded phases can be modelled by the use of the grafted chains with active groups.  Such chains are copolymers containing  a few  segments which very strongly attract fluid molecules.
We have carried out the calculations using the method described in reference \cite{c30}. We consider the grafted chains built of  $M_A$ segments $A$  and $M_B$ segments $B$ ($M_A+M_B=M$). The segments {A} mimic methylene groups in alkyl chains, while the segments $B$ correspond to the functional (active) group.
Within the framework of the model, the parameters characterizing the interactions of the $k$th component  with segments $A$ and $B$ are different ($\varepsilon^{(Ak)} \neq \varepsilon^{(Bk)}$).
Here  we show the results for the chains that contain $M_B=2$ segments of the $B$-type. These segments can be placed at consecutive positions of the backbone: $i_\text{s}$ and $i_{\text{s}+1}$.  The given type of chains is labeled as $Bi_\text{s}i_{s+1}$. We consider two positions of the functional group: in the close proximity to the wall (as the 3rd and 4th segments,  $B34$) and at the chain end ($B78$). The first type of the grafted chains corresponds to polar embedded stationary phases  which are often used in RPLC. The latter mimics stationary phases with functionalized terminal groups of the ligands \cite{c43}.  We have focused  only on the interactions of solvents with different segments. Therefore, we have assumed that both components are inert with respect to the wall. In the considered case, affinity of the solvent 1 to the segments $B$ is  much larger than the affinity to the `usual' segments $A$, $\varepsilon^{(A1)}=\varepsilon^{(\text{P}1)}=1.2$ and $\varepsilon^{(B1)}=3$. Moreover, $\varepsilon^{(AB)}=\varepsilon^{(\text{PP})}=1$ ($\chi^{(AB)}=0$). Interactions of the 2nd solvent with all segments are weaker, $\varepsilon^{(A2)}=\varepsilon^{(B2)}=\varepsilon^{(\text{P}2)}=1$.

\begin{figure}[!t]
\centerline{
\includegraphics[width=0.65\textwidth]{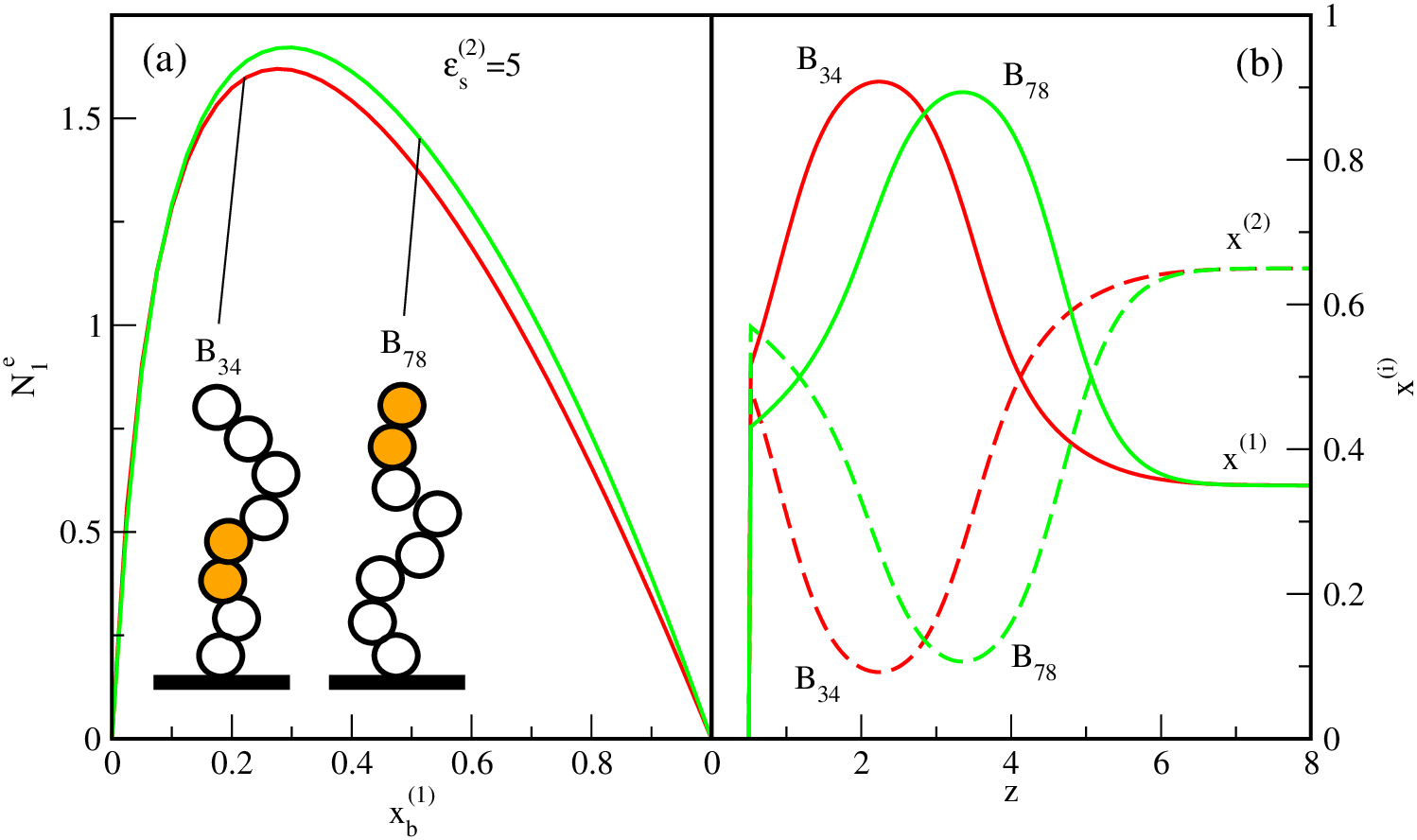}
}
\caption{(Color online) Relative excess adsorption isotherms (panel a) and  profiles of the solution composition at  $x^{(1)}_\text{b}=0.35$ (panel b) for two isomers of grafted chains  $B34$ and $B78$.  In panel b solid (dashed) lines correspond to the local mole fraction of the 1st (2nd) component. Parameters: $\rho_\text{P}=0.1$, $M=8$, $\varepsilon^{(\text{P}2)}=1$, $\varepsilon^{(A1)}=1.2$ and $\varepsilon^{(B1)}=3$.}
\label{fig:8}
\end{figure}

Figure~\ref{fig:8} illustrates the effect of the position of functional groups in  grafted chains on the relative excess adsorption isotherms and the composition of the surface layer. The shapes of the isotherm are typical of a strong preferential adsorption of the 1st component. The position of the functional group affects the relative excess adsorption for $x^{(1)}_\text{b}>0.1$.  Adsorption is higher for terminal functional groups. In panel b, the local mole fractions of the components are shown for $x^{(1)}_\text{b}=0.35$. There is a considerable enrichment of the  component 1  near the active groups. The maximum in $x^{(1)}$-profile is shifted toward the outer part of the surface layer for $B78$.

It follows from the above discussion that the adsorption of liquid mixtures on chemically bonded phases  and the composition of the surface layers depend on the relations between the parameters characterizing the system. The model is very sensitive to the choice of a particular set of the parameters. Changing these parameters one can simulate various systems.

Finally, we consider  the problem of `a range of the surface layer'. In chromatographic applications, a delimitation between the surface and bulk phases is necessary. The retention factor is proportional to the volume of the stationary (surface) phase. Moreover, the calculation of real adsorption (the number of molecules in the surface phase) requires  a definition of the boundary of the adsorbed phase.  In  interface science, the Gibbs dividing surface (GDS) is introduced to define the volumes of both phases.  The choice of the position of the Gibbs dividing surface is arbitrary \cite{c20}. When the density profiles are known, the GDS is usually defined via the standard equal area construction \cite{c44}. In analytical theories, the position of GDS is chosen using the procedure that ensures the thermodynamic consistency of the results \cite{c20}. Unfortunately, the problem is not trivial.

Various quantities can be treated as a measure of the thickness of the surface (stationary) phase. For example, one can use  characteristics of the chain layer: (i) the effective height of the grafted layer, $h_\text{eff}$ (the distance from the wall at which the segment density of the chains decreases to zero), and  (ii) the average brush height calculated from the following equation\cite{c45,c46}
\begin{equation} \label{eq:13}
h =2\frac{\int \rd z z\rho_\text{s}^{(\text{P})}(z)}{\int \rd z \rho_\text{s}^{(\text{P})}(z)}.
\end{equation}

The height of the polymer brush can quite well approximate the surface phase boundary since the partitioning is a dominant mechanism of the sorption.\cite{c5}  On the other hand, when adsorption plays a significant role in the process, the fluid density profiles should be analyzed to estimate: (i) the location of  the GDS \cite{c20} or (ii) the `effective' surface phase boundary. The latter is  the maximum distance, $z$,  above which the liquid is considered to be identical to the bulk solution \cite{c8}.

In all investigated systems, the bonded phases are highly inhomogeneous. One can divide the surface layer into three parts: (i) the interfacial region located near the wall, (ii) the interface between the brush and the bulk solution, and (iii) the middle (`bulk') part of the bonded phase. For relatively short grafted chains, the volume of the `bulk' stationary phase is comparable with the volumes of the interfacial regions.  There is no well-pronounced flat part in the segment density profiles (cf. figure~\ref{fig:1}) that could be treated as the `bulk' bonded phase. Therefore, the standard method for the location of the GDS \cite{c44} cannot be sufficiently precise. We assume that the boundary is located at the distance at which the composition of the adsorbed liquid becomes identical with the bulk solution (with a precision of $1$ percent of~$x_\text{b}^{(1)}$).

\begin{figure}[!t]
\centerline{
\includegraphics[width=0.5\textwidth]{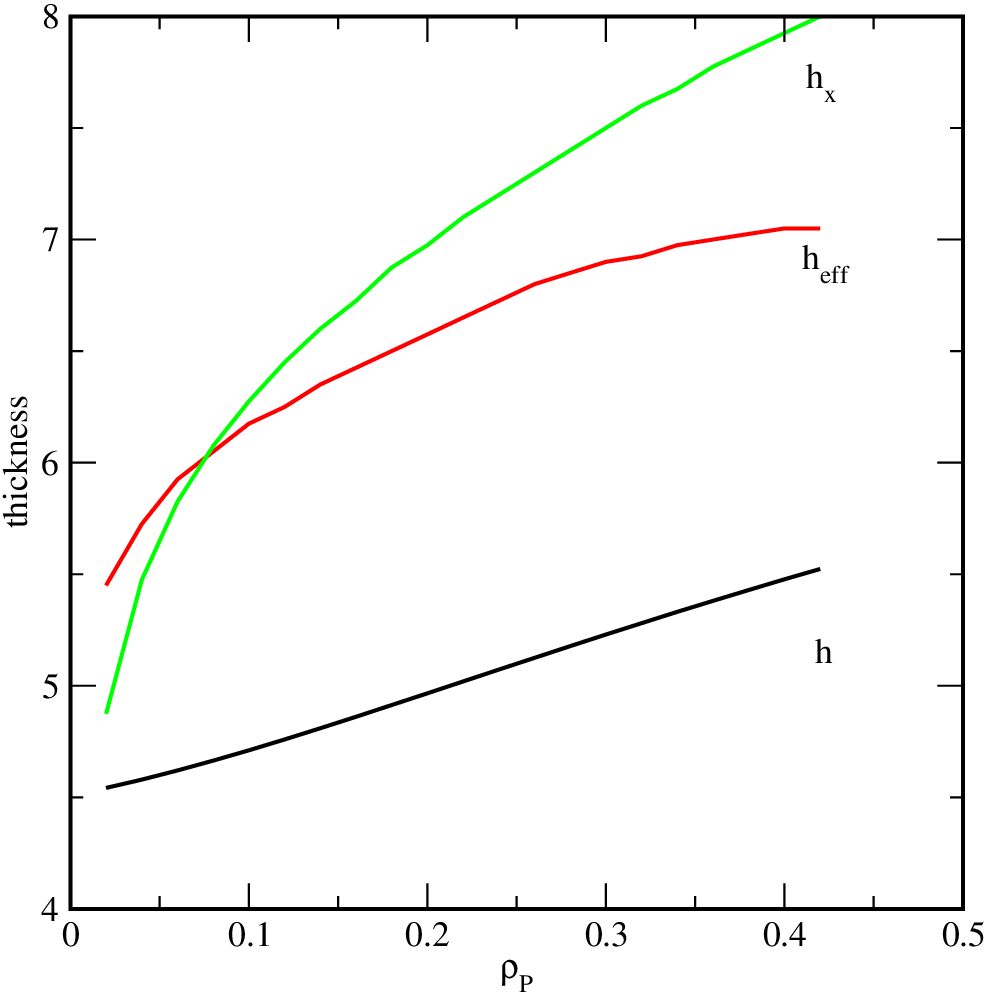}
}
\caption{(Color online) The effect of the grafting density on the average height of the chain layer, $h$, calculated from equation (\ref{eq:13}), on the effective thickness of the chain layer (the distance from the wall at which the segment density of the chains decreases to zero), $h_\text{eff}$, and on the  effective thickness of the surface layer (the maximum distance above which the liquid composition is the same as in the bulk solution), $h_x$,  at $x^{(1)}_\text{b}=0.35$. The remaining parameters are the same as those in figure~\ref{fig:1}.}
\label{fig:9}
\end{figure}

Figure~\ref{fig:9}  shows how the grafting density affects the average height of the chain layer $h$ [equation~(\ref{eq:13})], the `effective' height of polymer brush, $h_\text{eff}$, and the `effective' thickness of the surface layer, $h_x$. All the `thicknesses'  increase as the grafting density increases.  The dependence  of the average brush height  on the grafting density, $h$ vs $\rho_\text{P}$, has been the subject of numerous studies
\cite{c46,c47}.  The `effective' thickness of the brush is considerably  greater than the average height, $h_\text{eff}>h$ and $h_x>h$. As one can expected, the true adsorbed phase can be more expanded due to the secondary adsorption on the brush. Indeed, for $\rho_\text{P}>0.08$, the `effective' thickness of the surface phase  $h_x$  is considerably higher than $h_\text{eff}$.  This just reflects a competitive adsorption in the region above the chain layer. However, at low surface coverages, the opposite relation is found, $h_x<h_\text{eff}$. In the outer region of the brush, the density of chain segments is too low to change the liquid composition. After penetration of the solution into the chain layer, the composition starts to vary. The problem of the delimitation between  surface and bulk phases requires a further analysis.

The thickness of the surface layer varies with the change of the bulk solution. However,
these effects are not very significant \cite{c47}.

\clearpage

\section{Conclusions}

We have performed the density functional calculations to study the adsorption  from binary solutions on the chemically bonded phases. We have assumed that solvent molecules are spherical. The segments and the solvent molecules interact via Lennard-Jones (12--6) potential. All chain segments but the bonding segment are inert with respect to the substrate. The solvent molecules interacts with the solid surface via Lennard-Jones (9--3) potential or by the hard-wall potential.

We have systematically analyzed the effect of the selected factors on  the composition of the surface layer. In the model, both solvents are capable of penetrating the brush. Solvent molecules `stick' to the chains. Therefore, the solvent density profiles reflect the structure of the brush.  We have assumed that the solvent 1 exhibits high affinity to the grafted chains ($\varepsilon^{(\text{P}1)}>\varepsilon^{(\text{P}1)}$), while the solvent 2 very strongly interacts with the substrate ($\varepsilon^{(1)}_\text{s}<\varepsilon^{(2)}_\text{s}$). Molecules of different components compete for room inside the brush. As a consequence, the composition of the liquid mixture in the surface region changes with the distance from the wall.   A considerable enrichment of the 1st component was observed in the middle and in the outer parts of the surface region. On the contrary, close to the substrate, high concentration of the 2nd component was found.

We have discussed the effect of the system parameters on the profiles of local mole fractions of the components. We have studied the impact of the following parameters: the grafting density, the grafted chain length,  interactions of solvent molecules with grafted chains and the substrate and the presence of active groups in the chains. We have shown that the mole fraction of the 1st component increases as the grafting density increases. The same effect is observed as  the parameter $\varepsilon^{(\text{P}1)}$ rises or  the adsorption energy of the 2nd component, $\varepsilon^{(2)}_\text{s}$, decreases. The impact of the difference in interactions with the substrate is limited to its proximity while the effects of interactions with grafted chains are well pronounced  almost in the whole surface layer. An increase of the local mole fraction $x^{(1)}$ is also found for longer grafted chains.  The composition of the liquid mixture markedly changes near the active groups in the chains.

The local mole fractions of the liquid components inside the bonded phase cannot be experimentally measured. However, this is possible for the relative adsorption isotherms. Therefore, we have  presented the relative adsorption isotherms for the model systems considered.  We have qualitatively compared  our results with experimental isotherms. The theory well predicts  general trends observed in experiments~\cite{c20,c21,c22,c23,c24,c25,c26}.

We have also analyzed the problem of the delimitation between the surface phase and the bulk phase. The thickness of the surface phase is necessary for calculating the retention factors in chromatography.
We have compared three methods for the estimation of the thickness of the surface phase. The quantities characterizing the brush, namely, the average brush height, $h$ \cite{c45} or the effective brush thickness (the location of the brush edge), $h_\text{eff}$, can be effectively used  since the solvent molecules deeply penetrate the chain layer and do not accumulate `on the brush'. However, when the composition of the liquid mixture above the chain layer differs from that in the bulk solution, the effective surface layer thickness, $h_x$ should be estimated.

Our results confirmed that the density functional theory  provides a flexible and effective tool for modelling the  adsorption from binary solutions on the surfaces modified with grafted chains. The study of adsorption from binary solutions on the chemically bonded phase can be a starting point for the theoretical prediction of the solute retention from mixed mobile phases. The extension of  this approach is straightforward. Such a model system would contain at least four components: the bonded chains and three free components, i.e., two solvents and a solute. More realistic models  would be used to investigate chromatographic systems, e.g., the models involving associative and electrostatic interactions between components \cite{c30c,c30e}.

%
%

\clearpage

\ukrainianpart

\title{Адсорбція з бінарних розчинів на хімічно зв'язаних фазах}
\author{M. Борувко, T. Сташевский}
\address{Відділ моделювання фізико-хімічних процесів, Університет Марії Кюрі-Склодовської,
Люблін, Польща}

\makeukrtitle

\begin{abstract}
Ми використовуємо теорію функціоналу густини для дослідження адсорбції  рідких сумішей на поверхні твердого тіла, яка модифікована ланцюжками. Ланцюжки моделюються як вільно з'єднані сфери. Молекули рідини є сферичними. Взаємодія усіх типів сфер описується потенціалрм Ленарда-Джонса (12-6). Потенціал типу (9-3)  описує взаємодію молекул розчинника з підкладкою. Досліджено ізотерми надлишкової адсорбції, структуру поверхневого шару і його склад. Описано вплив густини і довжини ланцюжків, взаємодії молекул розчинника з ланцюжками і з підкладкою на адсорбцію. Розглянуто також вплив присутності активних груп
у ланцюжках.  Теоретичні результати узгоджуються з експериментальними спостереженнями.
\keywords теорія функціоналу густини, адсорбція з розчинів, полімер-модифіковані поверхні
\end{abstract}


\vspace{-3mm}

\begin{thebibliography}{99}

\bibitem{c1} Mittal V., In: Polymer Brushes: Substrates, Technologies and Properties,
Mittal~V.~(Ed.), CRC Press, Boca Raton, 2012, Chapter~1.

\bibitem{c2} Dorsey J.G., Dill K.A., Chem. Rev., 1989, \textbf{89}, 331;
\bibdoi{10.1021/cr00092a005}.

\bibitem{c3} Martire D.E., Boehm R.E., J. Phys. Chem., {1983}, \textbf{87}, 1045;
\bibdoi{10.1021/j100229a025}.

\bibitem{c4} Bor\'owko M., O\'s\'cik-Mendyk B., Bor\'owko P., J. Phys. Chem., 2005, \textbf{109}, 21056;
\bibdoi{10.1021/jp053195b}.

\bibitem{c5} Bohmer M.R., Koopal L.K., Tijssen R., J. Phys. Chem., 1991, \textbf{95}, 6285;
\bibdoi{10.1021/j100169a041}.

\bibitem{c6} Leermakers F.A.M., Philipsen  H.J.A., Klumperman B., J. Chromatogr. A, 2002, \textbf{959}, 37;
\\ \bibdoi{10.1016/S0021-9673(02)00382-5}.

\bibitem{c7} Bor\'owko M., R\.zysko W., Soko\l owski S., Staszewski T., J.  Phys.  Chem., 2009, \textbf{113}, 4763;
\bibdoi{10.1021/jp811143n}.

\bibitem{c8}  Bor\'owko M., Soko\l owski S., Staszewski T., J. Chromatogr. A, 2011, \textbf{1218}, 711;
\bibdoi{10.1016/j.chroma.2010.12.029}.

\bibitem{c9} Klatte S.J., Beck T.L., J. Chem. Phys., 1995, \textbf{99}, 16024;
\bibdoi{10.1021/j100043a049}.

\bibitem{c10} Rafferty J.L., Siepmann J.I., Schure M.R., J. Chromatogr. A, 2008, \textbf{1204}, 11;
\bibdoi{10.1016/j.chroma.2008.07.037}.

\bibitem{c11} Rafferty J.L., Siepmann J.I., Schure M.R., J. Chromatogr. A, 2008, \textbf{1204}, 20;
\bibdoi{10.1016/j.chroma.2008.07.038}.

\bibitem{c12} Rafferty J.L., Siepmann J.I., Schure M.R., Anal. Chem., 2008, \textbf{80}, 6214;
\bibdoi{10.1021/ac8005473}.

\bibitem{c13} Rafferty J.L., Siepmann J.I., Schure M.R., J. Chromatogr. A, 2009, \textbf{1216}, 2320;
\bibdoi{10.1016/j.chroma.2008.12.088}.

\bibitem{c14} Rafferty J.L., Siepmann J.I., Schure M.R., J. Chromatogr. A, 2012, \textbf{1223}, 24;
\bibdoi{10.1016/j.chroma.2011.11.039}.

\bibitem{c15}  Linsey R.K., Rafferty J.L., Eggimann B.L., Siepmann J.I., Schure M.R.,
J. Chromatogr. A, 2013, \textbf{1287}, 60;
\\ \bibdoi{10.1016/j.chroma.2013.02.040}.

\bibitem{c16} Fouqueau A., Meuwly M., Bermish R.J., J. Phys. Chem. B, 2007, \textbf{111}, 10208;
\bibdoi{10.1021/jp071721o}.

\bibitem{c17} Braun J., Fouqueau A., Bermish R.J., Meuwly M., Phys. Chem. Chem. Phys., 2008, \textbf{10}, 4765;
\bibdoi{10.1039/B807492E}.

\bibitem{c18} Dill K.A., J. Phys. Chem., 1987, \textbf{91}, 1980; \bibdoi{10.1021/j100291a060}.

\bibitem{c19}  Jaroniec M., Martire D.E., Bor\'owko M., Adv. Colloid Interface Sci., 1985, \textbf{22}, 177;
\\ \bibdoi{10.1016/0001-8686(85)80005-1}.

\bibitem{c20} Gritti F., Guiochon G., J. Chromatogr. A, 2007, \textbf{1155}, 85;
\bibdoi{10.1016/j.chroma.2007.04.024}.

\bibitem{c21} Gritti F., Kazakevich Y.V., Guiochon G., J. Chromatogr. A, 2007, \textbf{1169}, 111;
\bibdoi{10.1016/j.chroma.2007.08.071}.

\bibitem{c22} Kazakevich Y.V.,  LoBrutto R., Chan F., Patel T., J. Chromatogr. A, 2001, \textbf{913}, 75;
\bibdoi{10.1016/S0021-9673(00)01239-5}.

\bibitem{c23} Bocian S., Felinger A., Buszewski B., Chromatographia, 2008, \textbf{68}, S19;
\bibdoi{10.1365/s10337-008-0519-4}.

\bibitem{c24} Bocian S., Vajda P., Felinger A., Buszewski B., Anal. Chem., 2009, \textbf{81}, 6334;
\bibdoi{10.1021/ac9005759}.

\bibitem{c25} Bocian S., Vajda P., Felinger A., Buszewski B., Chromatographia, 2010, \textbf{71}, S5;
\bibdoi{10.1365/s10337-010-1522-0}.

\bibitem{c26} Buszewski B., Bocian S., Nowaczyk A., J. Sep. Sci., 2010, \textbf{33}, 2060;
\bibdoi{10.1002/jssc.201000101}.

\bibitem{c27} Bor\'owko M., Soko\l owski S., Staszewski T., J. Phys. Chem. B, 2012, \textbf{116}, 3115;
\bibdoi{10.1021/jp300114y}.

\bibitem{c28} Bor\'owko M., Soko\l owski S., Staszewski T., J. Phys. Chem. B, 2012, \textbf{116}, 12842;
\bibdoi{10.1021/jp305624n}.

\bibitem{c29} Bor\'owko M., Staszewski T., Condens. Matter Phys., 2012, \textbf{15}, 1;
\bibdoi{10.5488/CMP.15.23603}.

\bibitem{c30} Bor\'owko M., Soko\l owski S., Staszewski T., Mol. Phys., 2015, \textbf{113}, 1014;
\bibdoi{10.1080/00268976.2014.962636}.

\bibitem{c31} Yu Y.X., Wu J., J. Chem. Phys., 2002, \textbf{117}, 2368;
\bibdoi{10.1063/1.1491240}.

\bibitem{c32} Yu Y.X., Wu J., J. Chem. Phys., 2002, \textbf{117}, 10156;
\bibdoi{10.1063/1.1520530}.

\bibitem{c33} Yu Y.X., Wu J., J. Chem. Phys., 2003, \textbf{118}, 3835;
\bibdoi{10.1063/1.1539840}.

\bibitem{c30a} Bryk P., Soko\l owski S., J. Chem. Phys., 2004, \textbf{120}, 8299;
\bibdoi{10.1063/1.1695554}.

\bibitem{c30b} Bryk P., Soko\l owski S., J. Chem. Phys., 2004, \textbf{121}, 11314;
\bibdoi{10.1063/1.1814075}.

\bibitem{c30c} Bryk P., Pizio O., Soko\l owski S., J. Chem. Phys., 2005, \textbf{122}, 174906;
\bibdoi{10.1063/1.1888425}.

\bibitem{c30d} Bryk P., Pizio O., Soko\l owski S., J. Chem. Phys., 2005, \textbf{122}, 194904;
\bibdoi{10.1063/1.1898484}.

\bibitem{c30e} Tscheliessnig R., Billes W., Fischer J., Soko\l owski S., Pizio O., J. Chem. Phys., 2006, \textbf{124}, 164703;
\\ \bibdoi{10.1063/1.1898484}.

\bibitem{c30f} Bucior K., Fischer J., Patrykiejew A., Tscheliessnig R., Soko\l owski~S., J. Chem. Phys., 2007, \textbf{126}, 094704;
\\ \bibdoi{10.1063/1.2566372}.

\bibitem{c30g} Bor\'owko M., R\.zysko W., Soko\l owski S., Staszewski T., J. Chem. Phys., 2007, \textbf{126}, 214703;
\bibdoi{10.1063/1.2743399}.

\bibitem{c30i} Matusewicz M., Patrykiejew A., Soko\l owski S., Pizio O., J. Chem. Phys., 2007, \textbf{127}, 174707;
\bibdoi{10.1063/1.2780890}.

\bibitem{c30h} Patrykiejew A., Soko\l owski S., Tscheliessnig R., Fischer~J., Pizio~O., J. Phys. Chem. B, 2008, \textbf{112}, 4552;
\\ \bibdoi{10.1021/jp710978t}.

\bibitem{c34} Fundamentals of Inhomogeneous Fluids, Henderson D. (Ed.),
Marcel Dekker, New York, 1992.

\bibitem{c35} Rosenfeld J., Phys. Rev. Lett., 1989, \textbf{63}, 980;
\bibdoi{10.1103/PhysRevLett.63.980}.

\bibitem{c36} Wertheim M.S., J. Chem. Phys., 1987, \textbf{87}, 7323;
\bibdoi{10.1063/1.453326}.

\bibitem{c37} Bor\'owko M., Soko\l owski S., Staszewski T., J. Colloid Interface Sci., 2011, \textbf{356}, 267;
\bibdoi{10.1016/j.jcis.2011.01.023}.

\bibitem{c38} Weeks J.D., Chandler D., Andersen H.C., J. Chem. Phys., 1971, \textbf{54}, 5237;
\bibdoi{10.1063/1.1674820}.

\bibitem{c39} Everett D.H., Trans. Faraday Soc., 1965, \textbf{62}, 2478;

\bibitem{c40} Roe R.J., J. Chem. Phys., 1974, \textbf{60}, 4192;
\bibdoi{10.1063/1.1680888}.

\bibitem{c42} Bor\'owko M., In: Adsorption, Theory, Modeling and Analysis, Surfactant Science Series Vol.~107, T\'oth~J. (Ed.), Marcel Dekker, New York, 2002, Chapter~3.

\bibitem{c43} O'Sullivan G.P., Scully N.M., Glennon J.D., Anal. Lett., 2010, \textbf{43}, 1609;
\bibdoi{10.1080/00032711003653973}.

\bibitem{c44} Zangwill A., Physics at Surfaces, Cambridge University Press, Cambridge, 1988.

\bibitem{c45} Pastorino C., Binder K., Mueller M., Macromolecules, 2009, \textbf{42}, 401;
\bibdoi{10.1021/ma8015757}.

\bibitem{c46} Binder K., Milchev A., J. Polym. Sci., Part B: Polym. Phys., 2012, \textbf{50}, 1515;
\bibdoi{10.1002/polb.23168}.

\bibitem{c47} Bor\'owko M., Soko\l owski S., Staszewski T., J. Phys. Chem. B, 2013, \textbf{117}, 10293;
\bibdoi{10.1021/jp4027546}.

\end{thebibliography}
\end{document}